%% file: plos_gaps.tex
\documentclass[10pt]{article}

\usepackage{amsmath}
\usepackage{amssymb}

\usepackage{graphicx}

\usepackage{cite}

\usepackage{color} 
\usepackage{comment}

\usepackage[labelfont=bf,labelsep=period,justification=raggedright]{caption}

\makeatletter
\renewcommand{\@biblabel}[1]{\quad#1.}
\makeatother

\date{}

\begin{document}

\begin{flushleft}
{\Large
\textbf{Improving contact prediction along three dimensions}
}
\\
Christoph Feinauer$^{1,\dagger}$, 
Marcin J. Skwark$^{3,4,\dagger}$, 
Andrea Pagnani $^{1,2}$,
Erik Aurell$^{3,4,5,\ast}$
\\
\bf{1} DISAT and Center for Computational Sciences, Politecnico Torino, Corso Duca degli Abruzzi 24, I-10129 Torino, Italy
\\
\bf{2} Human Genetics Foundation-Torino, Molecular Biotechnology Center, Via Nizza 52, I-10126 Torino, Italy
\\
\bf{3} Department of Information and Computer Science, Aalto University, P.O. Box 15400, FI-00076 Aalto, Finland
\\
\bf{4} Aalto Science Institute (AScI), Aalto University, PO Box  15600, FI-00076 Aalto, Finland
\\
\bf{5} Dept.~Computational Biology, AlbaNova University Centre, SE-106 91 Stockholm, Sweden
\\
$\ast$ E-mail: eaurell@kth.se
\\
$\dagger$ These authors contributed equally to this work
\end{flushleft}

\section*{Abstract}
\input{abstract.tex}
\section*{Author Summary}
\input{author-summary.tex}

\include{introduction}
\include{results}

\include{discussion}

\include{methods}

\section*{Acknowledgments}
CF, MS and EA thank Magnus Ekeberg and Tuomo Hartonen for 
valuable discussions. 
This work has been supported by the People Programme
(Marie Curie Actions) of the European Union’s Seventh Framework 
Programme FP7/2007-2013/ under REA grant agreement n. 290038 (CF) and
by the Academy of Finland through the Academy of Finland Center of Excellence COIN (MS and EA). 
We acknowledge the computational resources provided by Aalto Science-IT project. 

\bibliography{bibliography,bibliography-AP2,bibliography-EA}

\section*{Figure Legends}

\section*{Tables}

\section*{Supporting Information Legends}

\paragraph{Supporting information S1.} Figures and tables in this supplementary material are numbered
identically as in the main paper, and are based on the same data, the only difference being that we use throughout
the $8.5$\AA\ heavy atom criterion, which has been used in the previous work~\cite{ekeberg_improved_2013,ekeberg2014}

\end{document}

%% file: abstract.tex
Correlation patterns in multiple sequence alignments of homologous proteins can be
exploited to infer information on the three-dimensional structure of their
members. The typical pipeline to address
this task, which we in this paper refer to as
the \textit{three dimensions of contact prediction},
is to (i) filter and align the raw
sequence data representing the evolutionarily related proteins; (ii) choose a predictive model to describe a 
sequence alignment; 
(iii) infer the model parameters and interpret them 
in terms of structural properties, such as an accurate contact map.
We show here that all three dimensions are important for overall
prediction success. In particular, we show that
it is possible to improve significantly along the second dimension by going beyond the pair-wise Potts 
models from statistical physics, which have hitherto been the focus of
the field. These (simple) extensions are motivated
by multiple sequence alignments often containing long stretches of gaps 
which, as a data feature, would be rather untypical for
independent samples drawn from a Potts model.
Using a large test set of proteins we show that the combined
improvements along the three dimensions are as large as any reported to date.

%% file: author-summary.tex
Proteins are large molecules that living cells make by stringing together
building blocks called amino acids or peptides, following their blue-prints in the
DNA. Freshly made proteins are typically long, structure-less chains of peptides,
but shortly afterwards most of them fold into 
characteristic structures.
Proteins execute many functions in the cell, for which they need to have
the right structure, which is therefore
very important in determining what the proteins can do.

The structure of a protein can be determined by X-ray diffraction and other
experimental approaches which are all, to this day, somewhat labor-intensive
and difficult. On the other hand, the order of the peptides in a protein
can be read off from the DNA blue-print, and such protein sequences are today
routinely produced in large numbers. 

In this paper we show that many similar protein sequences can be used to 
find information about the structure. The basic approach is to construct a 
probabilistic model for sequence variability, and then to use the parameters 
of that model to predict structure in three-dimensional space. 
The main technical novelty compared to previous
contributions in the same general direction is that we use models more
directly matched to the data.

%% file: introduction.tex
\section*{Introduction}

The large majority of cellular mechanisms are executed and controlled
by the coordinated action of thousands of proteins, whose biological
function is strongly connected to their three-dimensional (3D)
arrangement. As shown by Anfinsen almost 40 years ago
\cite{Anfinsen1973}, the native three-dimensional structure and
function of any given protein is unambiguously encoded by its amino acid
sequence. Despite many years of intensive work in the field, and many
partial successes, the problem of predicting structural
properties of a protein from sequence information alone is still to
be considered as an open problem.

Recent years have seen a staggering increase in the amount of available protein sequence data,
which can be attributed to the developments in the sequencing technologies. Currently, more than
52 million protein sequences are known, which is a figure that continues growing by over 50\% yearly~\cite{uniprot2013update}.
This, coupled with advances in sequence homology detection methods~\cite{punta2012,Hmmer2011,Hhblits2011}, allows for construction
of accurate multiple sequence alignments (MSA), capable of capturing the evolutionary history of proteins of interest.
As a result of the tradeoff between the evolutionary drift and the constraint imposed by biological function, proteins
comprising such a multiple sequence alignment are generally characterized by: (i) a considerable sequence
variation, (ii) a striking similarity between their 3D structures.
In particular, the evolutionary pressure to conserve structure
suggests that residues in spatial proximity should exhibit patterns of
correlated amino-acid substitutions in these multiple sequence alignments.

The approach of using co-evolutionary information encoded in the MSA of homologous proteins
to predict structural features of its members was proposed
long ago
\cite{Altschuh1987,gobel_correlated_1994,Neher1994,Shindyalov1994,Lockless1999,Fodor2004}
(see also \cite{marks2012, valencia2013} for recent reviews on the subject).
The last five years have witnessed a renewed interest in the problem: after a
first wave of works inspired by statistical physics based on Bayesian methods
\cite{Burger_2007, burger_disentangling_2010}, or on different mean-field
approximations to a maximum-entropy model
\cite{weigt_identification_2009,morcos_direct-coupling_2011}, a burst of
scientific activity produced new and increasingly accurate global inference
methods \cite{balakrishnan2011,Dijk2011,jones2012,cocco_principal_2013,ekeberg_improved_2013,kamisetty_assessing_2013,ekeberg2014}. 
Apart from inferring structural properties for
single protein domains, co-evolutionary methods provide reliable predictions
for: (i) inter-chain structural organization
\cite{morcos_direct-coupling_2011}, (ii) specificity and partner identification
in protein-protein interaction in bacterial signal transduction system
\cite{burger_disentangling_2010,Procaccini2011}, (iii) essential
residue-residue contacts to determine native 3D structures
\cite{marks_protein_2011,hopf2012,Sulkowska2012}.

The basis of all these computational methods is the idea of global statistical
inference. The global approach has the advantage that it is able to
disentangle direct from indirect couplings between residues. By modeling the
whole data set at once, and not only pairs of residues independently, it is, for
example, possible to identify a case in which high correlation between two
residues is the indirect consequence of both being directly correlated to a
third variable.

Methods that address this problem are collected under the umbrella term of
\emph{Direct Coupling Analysis} (DCA).  Some methods used so far are (i) the
message passing based DCA (mpDCA) \cite{weigt_identification_2009} and the
mean-field DCA (mfDCA) \cite{morcos_direct-coupling_2011}, (ii) sparse
inverse covariance methods (PSICOV) \cite{jones2012}, (iii) pseudo-likelihood
based optimization \cite{balakrishnan2011,ekeberg_improved_2013,kamisetty_assessing_2013}. 
The techniques proposed in (iii), and in particular
the {\em plmDCA} algorithm \cite{ekeberg_improved_2013,ekeberg2014}, seem to
achieve the most accurate predictions so far, when validated against
experimentally determined protein structures. Nonetheless, plmDCA shows
systematic errors that can be traced back to certain intrinsic characteristics
of MSAs, such as the existence of repeated gap stretches in specific parts of
the alignment. This phenomenon reflects the tendency of homologous proteins to
include large-scale modular gene rearrangements in their phylogenetic
evolution, as well as point insertions/deletions. As an empirical way to
describe such complex rearrangements, sequence alignment methods 
typically use a form of substitution matrix to assign scores to amino acid
matches and a gap penalty for matching an amino-acid in one sequence and a gap
in the other. In either case, the most widely utilized gap-penalty schemes
assign a large cost to open a gap and a smaller one to extend a gap, so that
the overall penalty $Q$ of creating a stretch of gaps of length $l$ is $Q(l) =
a + b (l - 1)$, where typically $a \sim -10$ and $b \sim
-2$~\cite{Durbin_Book1998}. This introduces an intrinsic asymmetry between gaps
and amino acids, where subsequences consisting only of the gap variable are much
more likely to occur in an MSA than subsequences of one and the same amino
acid.  

In this work we highlight that contact prediction can be improved in three
different ways, or \emph{dimensions}, all important for overall success and
accuracy.  The first dimension is \textbf{Data}; it matters which MSA one
uses as input to a DCA scheme.  Continuing recent work of one of
us~\cite{skwark2013pconsc} we show that in a large test data set MSAs built on
HHblits alignments give more useful information than MSAs derived from the Pfam protein
families database. This conclusion is perhaps not surprising, as the
Pfam database was not constructed with potential applications to DCA in mind,
but is practically important if DCA is to reach its full potential.  The second
dimension is \textbf{Model}; it matters which global model one tries to learn
from an MSA, and it is possible to systematically improve upon the pairwise
interaction models, or Potts models, which have hitherto been the focus of the
field. This we show starting from the empirical observation that several DCA
methods typically produce high-ranking false positives in parts of an alignment
rich in gaps, and the simple fact that any subsequence of one of the same
variable has low sequence entropy, and is thus unlikely to occur in random
samples drawn from a Potts model, unless its model parameters take special
values, \textit{i.e.} unless at least some of them are quite large.  We
therefore enhance the Potts model by including terms depending on gaps of any
length, much in the spirit of a simplified model for protein folding proposed
long ago \cite{Wako_Sato_1978}. In this way we are able to effectively reduce
the false positive rate in gap-rich regions of the MSA over a large test data
set of diverse proteins.  The third dimension is \textbf{Method}. It is well
known that DCA by learning a Potts model describing an MSA by exactly
maximizing a likelihood function is computationally unfeasible for realistic
protein sizes.  Most DCA methods can therefore be seen as circumventing this
fact, either by approximating the likelihood function, or by using a different
(weaker) learning criterion. Here, we show that pseudo-likelihood based
optimization methods, which have demonstrated the best performance among standalone
methods, have the additional advantage of being flexible and easily adaptable
to learning other \textit{models}. This we show by including terms depending on
gaps of any length in the score function optimized in the recently developed
\textit{asymmetric} version of the plmDCA algorithm~\cite{ekeberg_improved_2013,ekeberg2014}.

Important recent developments, not touched upon in the present work, are combining two or more DCA methods
and/or incorporating supplementary information in a prediction process, as done in~\cite{skwark2013pconsc} 
and~\cite{kamisetty_assessing_2013}. One motivation is that it is theoretically interesting by itself
to see how much useful information can be learned by simply starting from 
the data, proposing a model, and then learning the model more or less well from the data; 
a second motivation is computational speed, as a stand-alone method is (typically) much faster than meta-predictors.
A pragmatic motivation for this choice is that any meta-predictor
is based on combining stand-alone methods. Hence, improving stand-alone methods gives
scope for further improvements of the meta-predictors. Indeed, we believe that the method
developed here should allow for further improvements to the methods of~\cite{skwark2013pconsc} 
and~\cite{kamisetty_assessing_2013}; this we leave however for future work.

%% file: results.tex
\section*{Results}

\paragraph{We have developed a new fast DCA method by extending the Potts model with gap parameters.}
The new method \textit{gap-enhanced pseudo maximim-likelihood direct contact analysis} (gplmDCA)
uses as underlying inference engine the recent \textit{asymmetric pseudo maximum-likelihood}~\cite{ekeberg2014}
augmented by gap parameters, as described in Methods.
The added gap parameters have the same status as the other parameters of the model, and the inference task posed by
gplmDCA is therefore formally the same as in plmDCA. The number of additional
parameters is less than $\frac{N^2}{2}$, with $N$ being the length of a
alignment, a small fraction of the number of parameters in Potts model based DCA. We have found that the computing time
our new method gplmDCA is almost indistinguishable from the asymmetric version 
of plmDCA~\cite{ekeberg2014}.

\begin{figure}[h!]
\begin{center}
\includegraphics[width=0.45\textwidth]{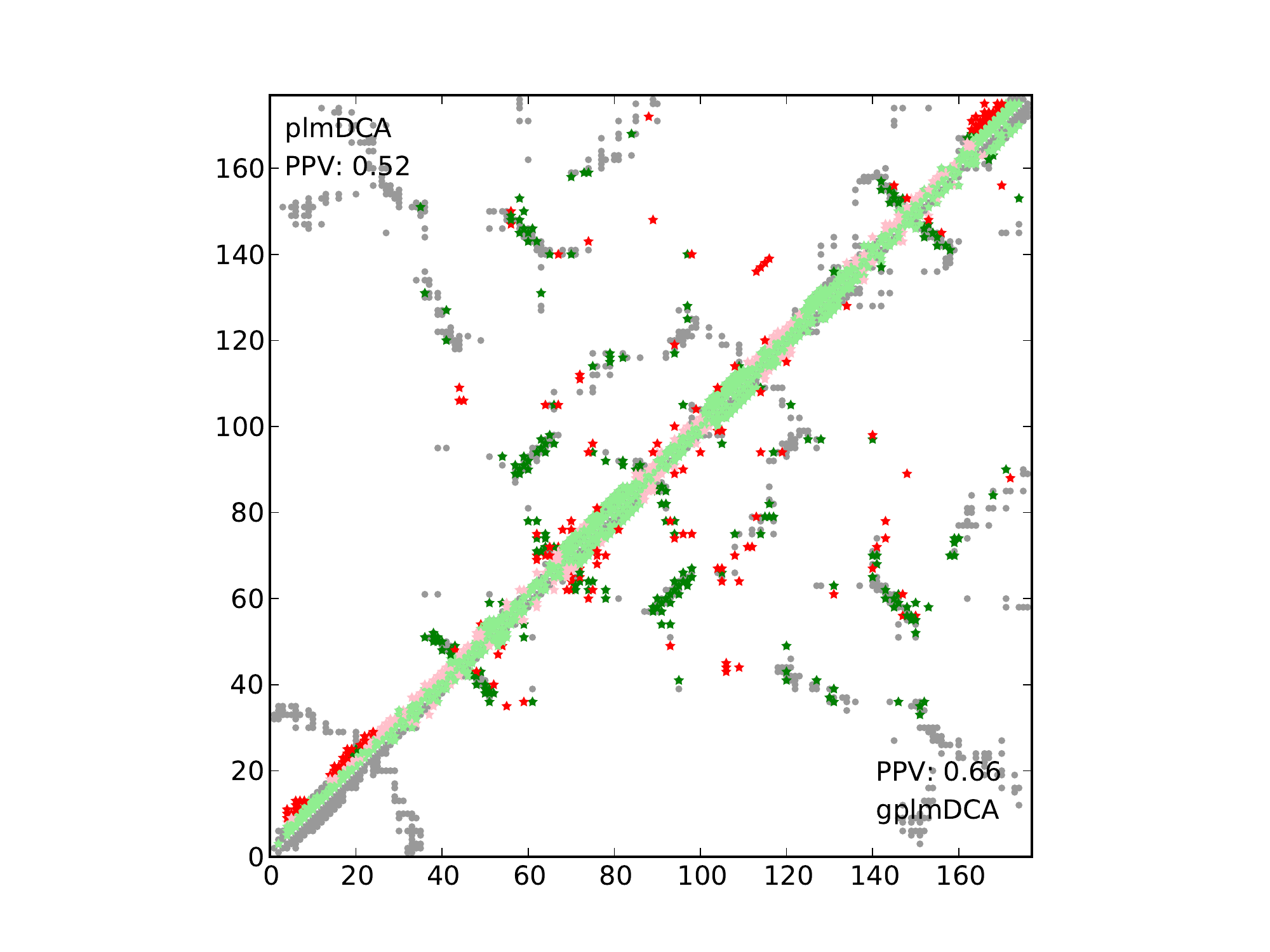}
\includegraphics[width=0.45\textwidth]{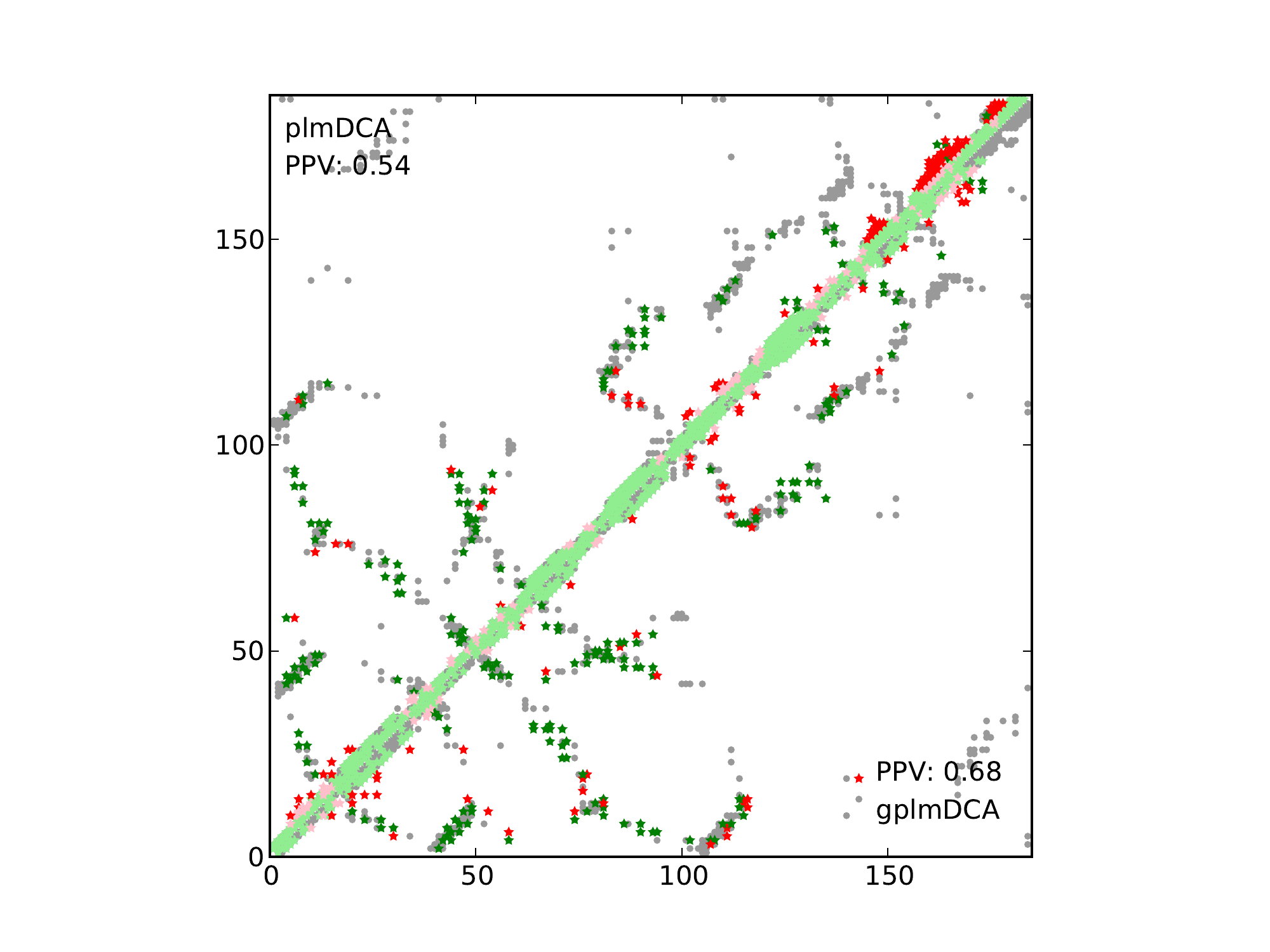}
\end{center}
\caption{Examples of qualitative contact prediction improvement. Gray circles:
contacts observed in crystal structure, Stars: predicted contacts (green:
correctly predicted, right: incorrectly predicted). Predicted short-range
contacts (not considered in the assessment) are drawn in pale colors. Left
panel: contact prediction maps built by plmDCA and gplmCDA using protein
sequences homologous to 1JFU:A as explained in Methods.  plmDCA here predicts a
number of strong couplings at both the N-terminus and the C-terminus which
arise from the high sequence variability at both ends of proteins homologous to
1JFU:A, and the many gaps in the multiple sequence alignments at these
positions. In gplmDCA these gaps lead to adjustment of gap parameters and not
to contact predictions. Right panel: analogous results using protein sequences
homologous to 1ATZ where gplmDCA removes strong spurious couplings at the
C-terminus.  } 
\label{example-good} 
\end{figure}

This introduction of gap parameters significantly alleviates a well-known negative 
trait of plmDCA -- the presence of gap-induced artifacts in many contact maps. 
The reduction of strong, but spurious couplings in the inference process allows for the detection
of other couplings, improving prediction qualitatively. 
Figure~\ref{example-good} shows two examples where conspicuous incorrect predictions 
at the N-terminus and the C-terminus are removed.

\begin{figure}[h!]
\begin{center}
\includegraphics[width=0.44\textwidth]{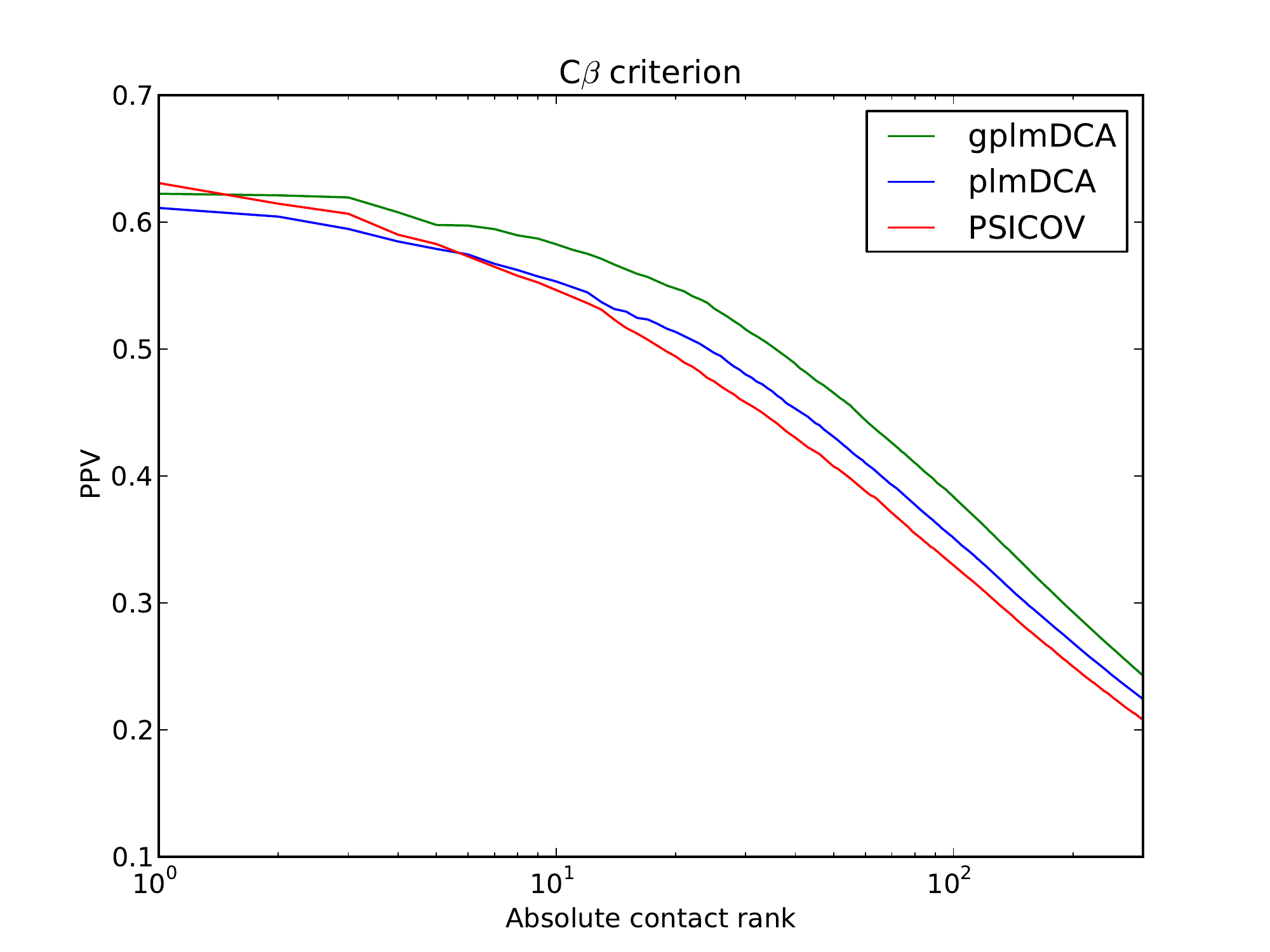}
\includegraphics[width=0.44\textwidth]{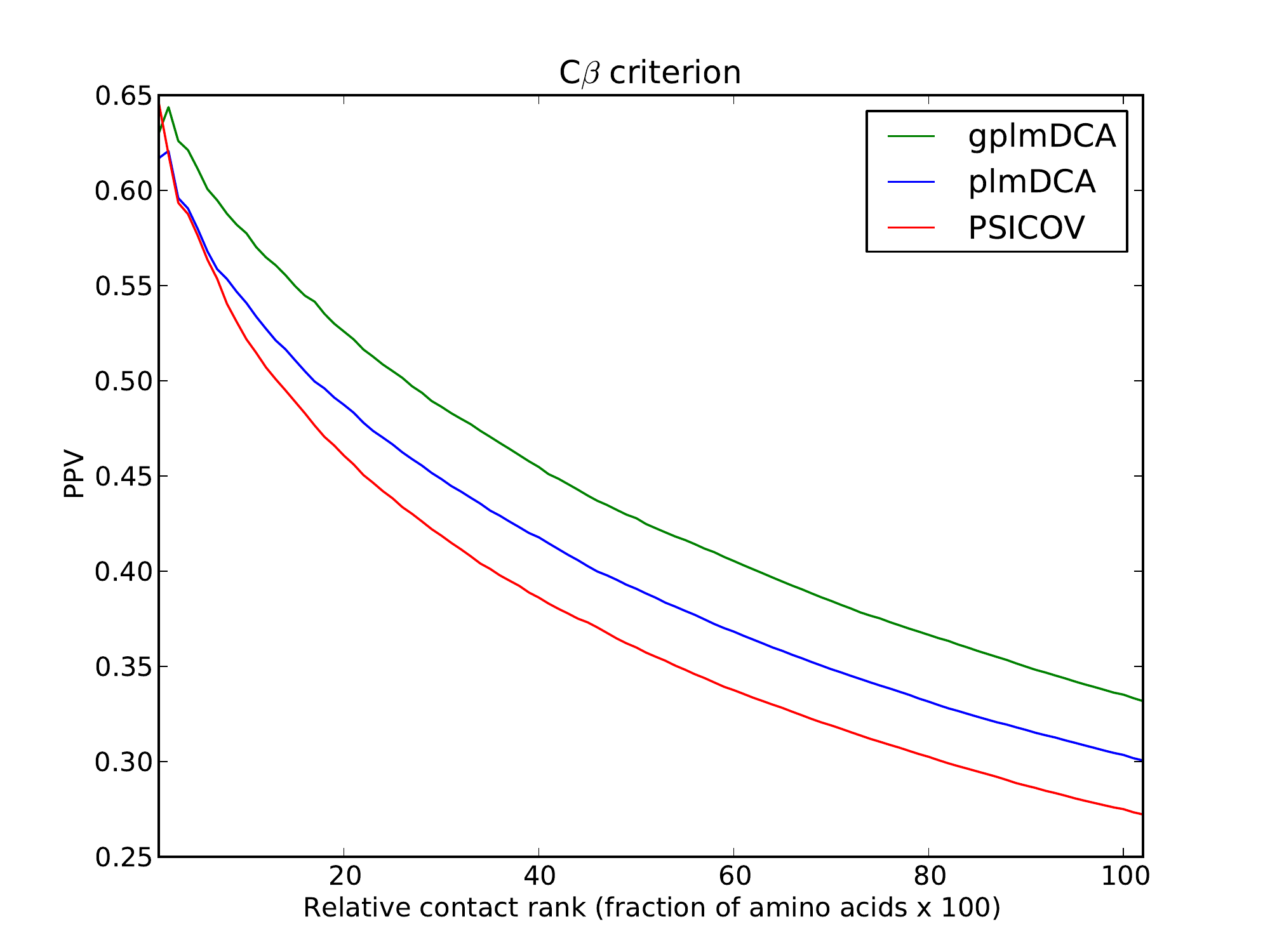}
\end{center}
\caption{Prediction precision (PPV), average over all proteins in the main test data set. The curves show for PSICOV, plmDCA and gplmDCA
the average of the number of correct predictions in the $n$ highest scoring pairs divided by $n$.  Left panel: PPV for absolute contact index; the horizontal axis shows $n$.
gplmDCA yields higher absolute PPV than plmDCA for all $n$. PSICOV is more often right than either plmDCA or gplmDCA in its prediction
of the very first (strongest) contact ($n=1$), but is inferior to plmDCA at $n=5$ and larger, for this test set. Right panel: PPV for relative contact index (fraction of protein length). the horizontal axis shows $(n/N) \cdot 100$.
In contrast to the curves of absolute PPV in left panel, the relative order of the three methods is
here uniform in $n$ reflecting that the advantage of PSICOV for the first contact is weaker for
longer proteins, which matter more at the ordinate of the graph of relative PPV. 
}
\label{ppv-plots}
\end{figure}

\paragraph{Adding gap parameters to the model improves contact predictions overall.}
Using a large test set, the \textit{main data set} as described in Methods, we have found that adding gap parameters
increases positive predictive value (PPV) for a large majority of all proteins in the data set.
This increase holds for our main criterion ($C\beta$ criterion)
for both absolute PPV and PPV relative to protein length, see Figure~\ref{ppv-plots}.
The average relative improvement of gplmDCA over plmDCA, as measured by mean absolute PPV,  
is 16.7\% (6.7\% to 26.7\% within a 95\% confidence interval).
In this paper our focus is on the possibility of learning models which lead to better
contact prediction, and not of learning a given model more or less well.

To set a scale of the improvement we include however in the comparisons in Figures~\ref{ppv-plots}
also PSICOV~\cite{jones2012}, another leading approach to the DCA which can be understood to learn
the same model as plmDCA but by a different inference method.  

Supplementary material contains results of the analysis conducted in this
paper based on our former criterion (8.5 \AA{} heavy atom criterion) for the
sake of immediate backwards comparability with previous
work~\cite{ekeberg_improved_2013,ekeberg2014}.

\begin{figure}
\begin{center}
\includegraphics[width=0.7\textwidth]{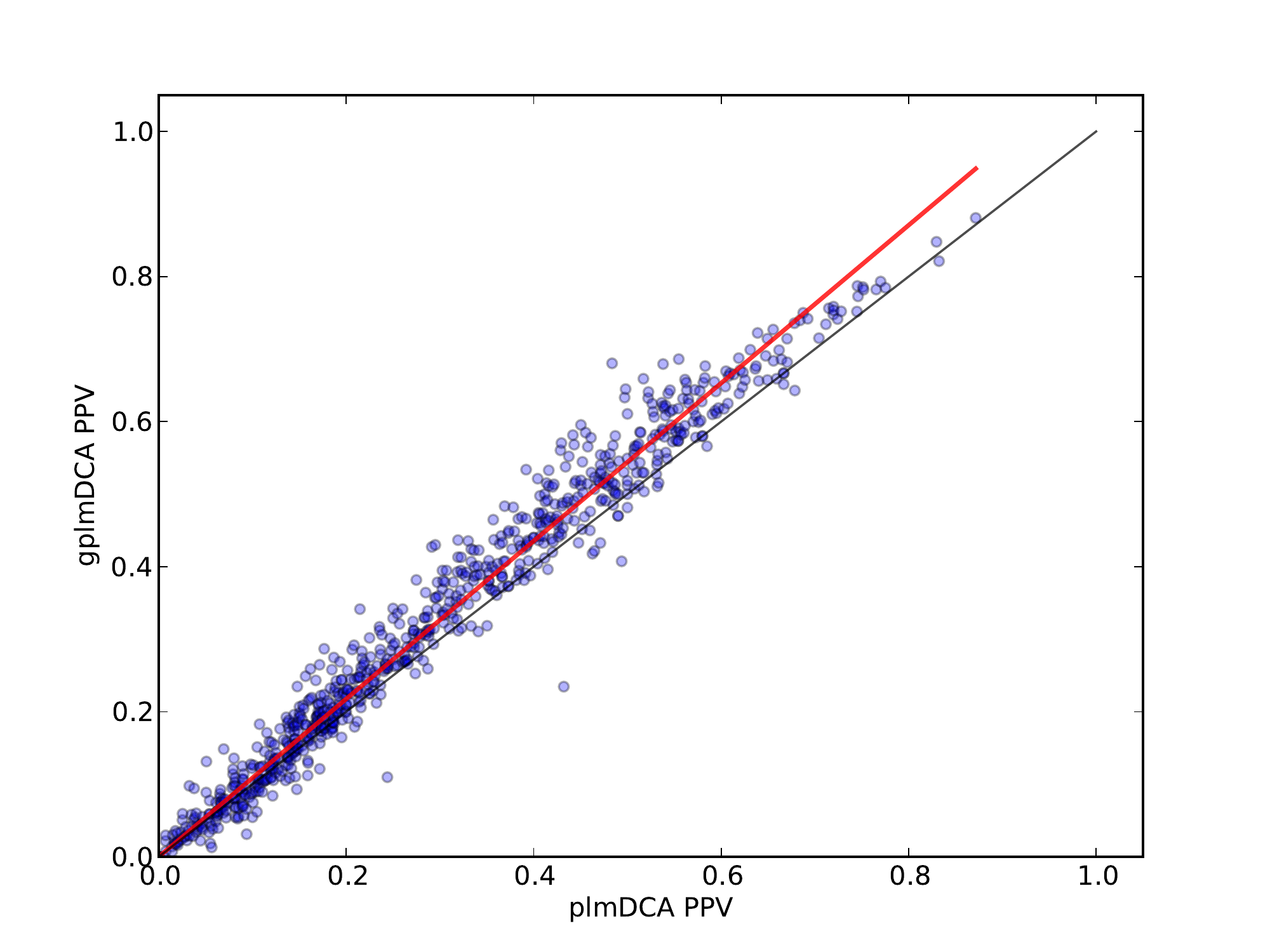}
\end{center}
\caption{Contact prediction accuracy (mean absolute PPV)
for proteins in the main test set by plmDCA (abscissa) and gplmDCA (ordinate). 
Most of the points fall above the diagonal indicating that gplmDCA is more accurate 
than plmDCA for most of proteins in the test set. 
Data points can be fitted a straight line by Ordinary Least Squares regression, with slope $1.0885 \pm 0.004$ ($R^2=0.992$) 
indicating that gplmDCA is generally relatively more accurate than plmDCA the more
accurate is plmDCA itself. There are two noticeable negative outlier counter-examples,
predictions for proteins with PDB identifiers 1CXY:A and 3P8B:A, with plmDCA and gplmDCA PPV scores of 0.43 vs 0.23 and 0.24 vs 0.10, respectively. Both are analysed in Discussion.
}

\label{scatter-plot}
\end{figure}

\paragraph{Adding gap parameters to the model improves individual contact predictions.}
A regression analysis of prediction accuracy, as measured by absolute PPV, reveals
clear systematic differences between plmDCA and gplmDCA.
As shown in Figure~\ref{scatter-plot} the overall advantage of gplmDCA primarily arises
from proteins where PPV is relatively high, \textit{i.e.} where prediction 
by plmDCA itself is accurate.

Quantitative statistics of this effect are summarized in Table~\ref{ppv-table}.
Including all 801 proteins in the main test set we find that in 84\% of the cases gplmDCA
does at least as well as plmDCA, but if we include only the 665 instances where the PPV
from both plmDCA and gplmDCA are larger than a relatively low cut-off of $0.1$ 
this fraction rises to 89\%, eventually reaching 95\%.
\begin{table}[h!]
\centering
\begin{tabular}{|l | l | l l|}
\hline
Cutoff & Proteins & Better & Better or equal\\
\hline
0.50 &  148 &  132 (0.89) &  141 (0.95) \\
0.40 &  258 &  227 (0.88) &  241 (0.93) \\
0.30 &  358 &  315 (0.88) &  334 (0.93) \\
0.20 &  478 &  417 (0.87) &  447 (0.94) \\
0.15 &  575 &  489 (0.85) &  530 (0.92) \\
0.10 &  665 &  536 (0.81) &  593 (0.89) \\
0.05 &  745 &  567 (0.76) &  639 (0.86) \\
\hline
ALL  &  801 &  587 (0.73) &  676 (0.84) \\
\hline
\end{tabular}
\caption{Numbers and fraction of proteins where gplmDCA performs better than plmDCA.
In each row all proteins in the data set are included for which the PPV from both plmDCA
and gplmDCA is larger than the cutoff value given in the first column.
The full data set (last row) consists of 801 proteins for 587 (73\%) of which gplmDCA 
performs better than plmDCA. In the most stringent selection (first row) there are
148 proteins where both plmDCA and gplmDCA have a PPV of at least 0.5. In this set
gplmDCA performs better on 132 (89\%) of the instances.
}
\label{ppv-table}
\end{table}

It is evident that the expected utility of DCA-like contact prediction is heavily dependent on the information
content in the input alignment. The information content is closely correlated to the number of unique protein
sequences in the alignment. Until recently, it has been a rule of thumb that one needs at least 10 times as many sufficiently
diverse proteins in the alignment as there are amino acids in the protein in question. That meant that contact prediction
with alignments of fewer than 1000 sequences was considered unfeasible.

\begin{figure}[h!]
\begin{center}
\includegraphics[width=0.44\textwidth]{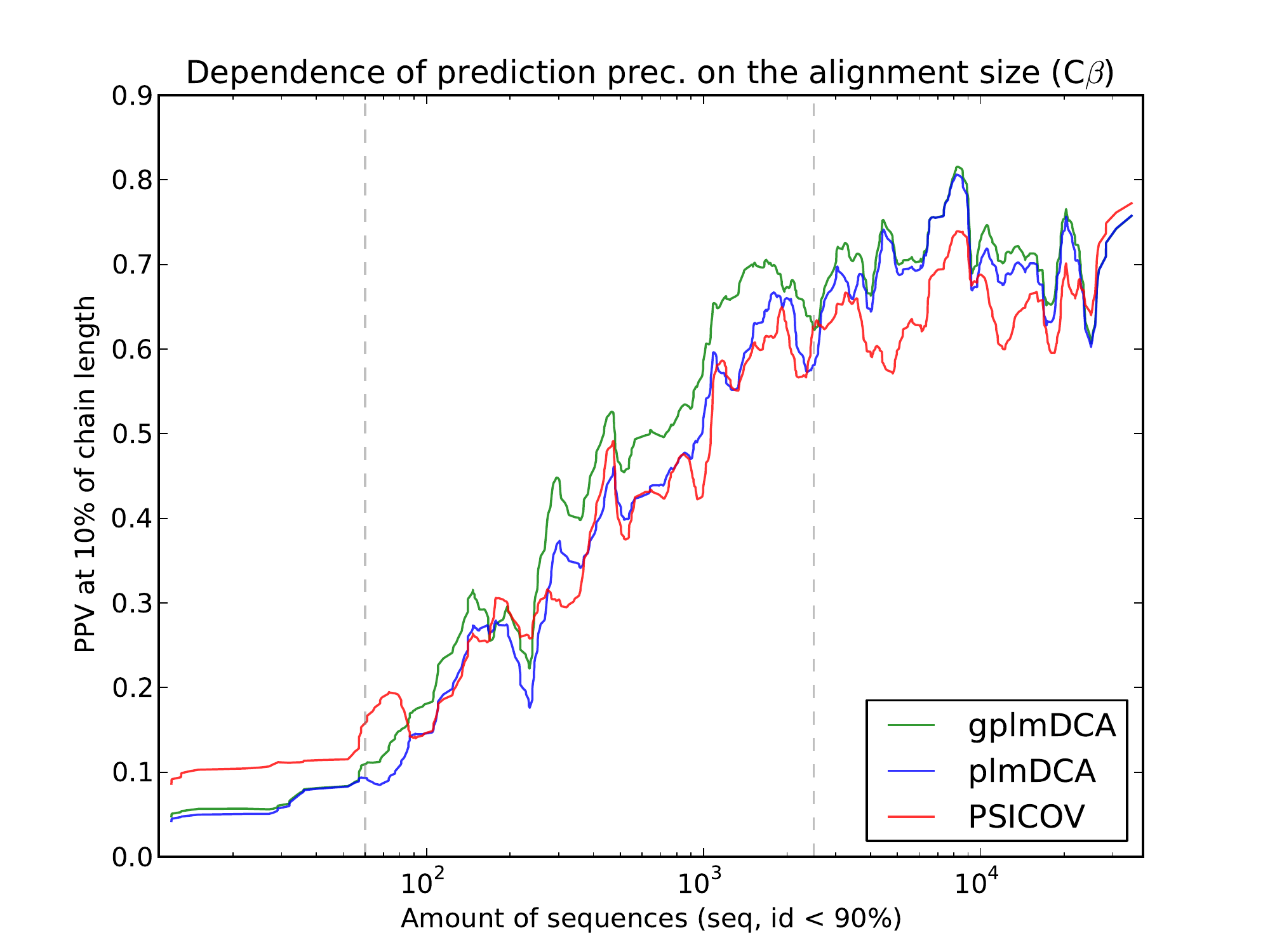}
\hfill
\includegraphics[width=0.44\textwidth]{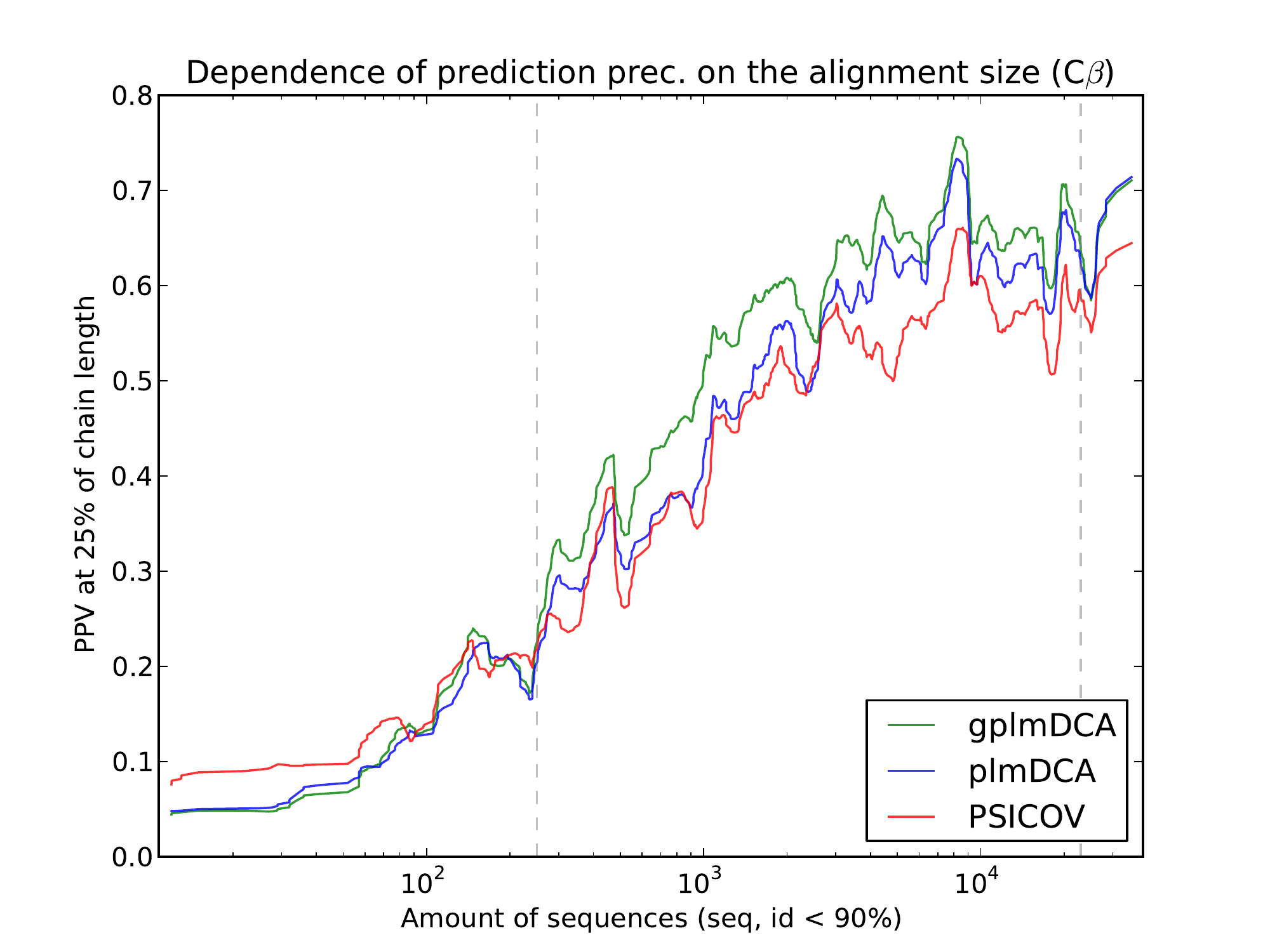}

\includegraphics[width=0.44\textwidth]{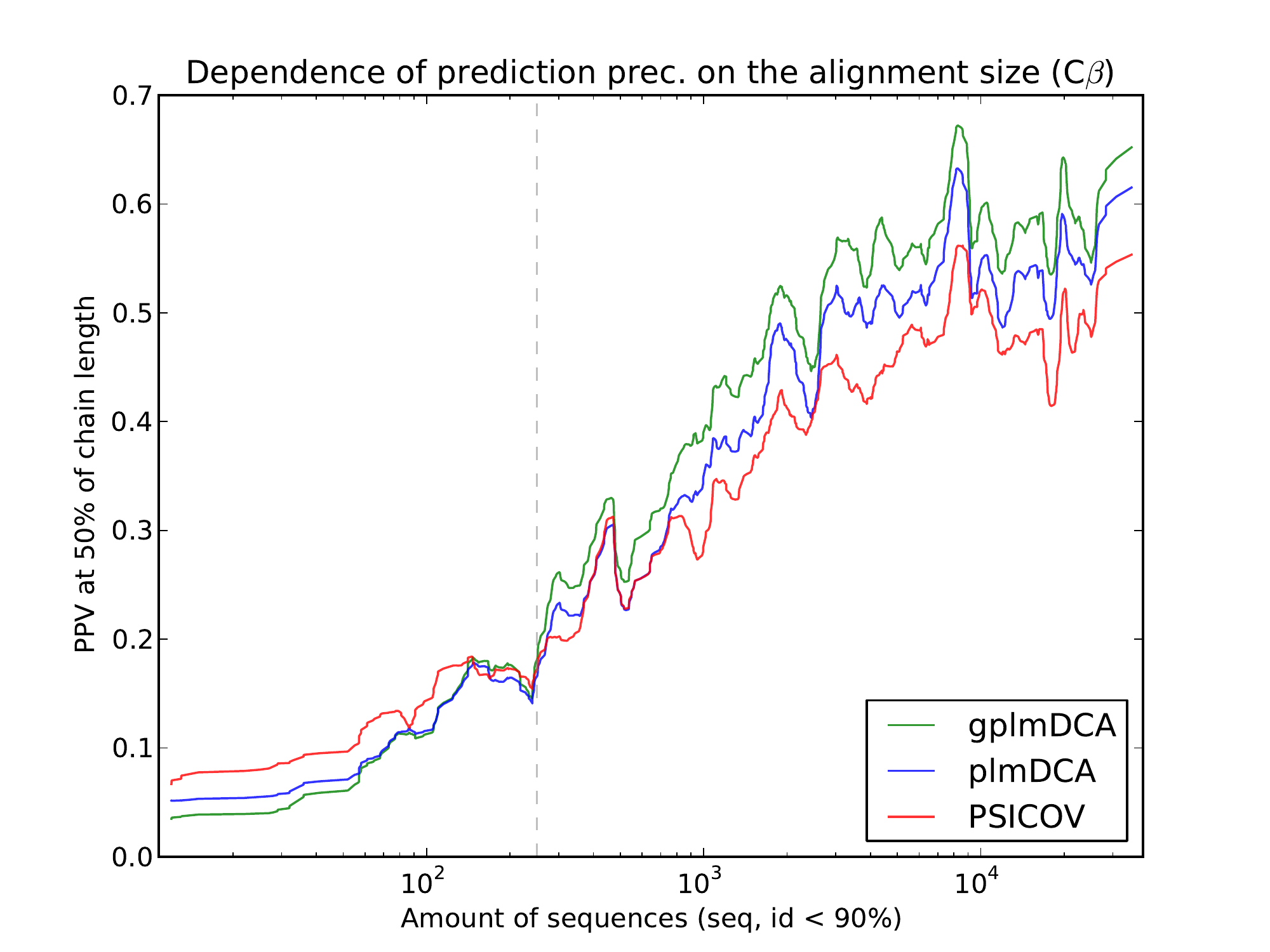}
\hfill
\includegraphics[width=0.44\textwidth]{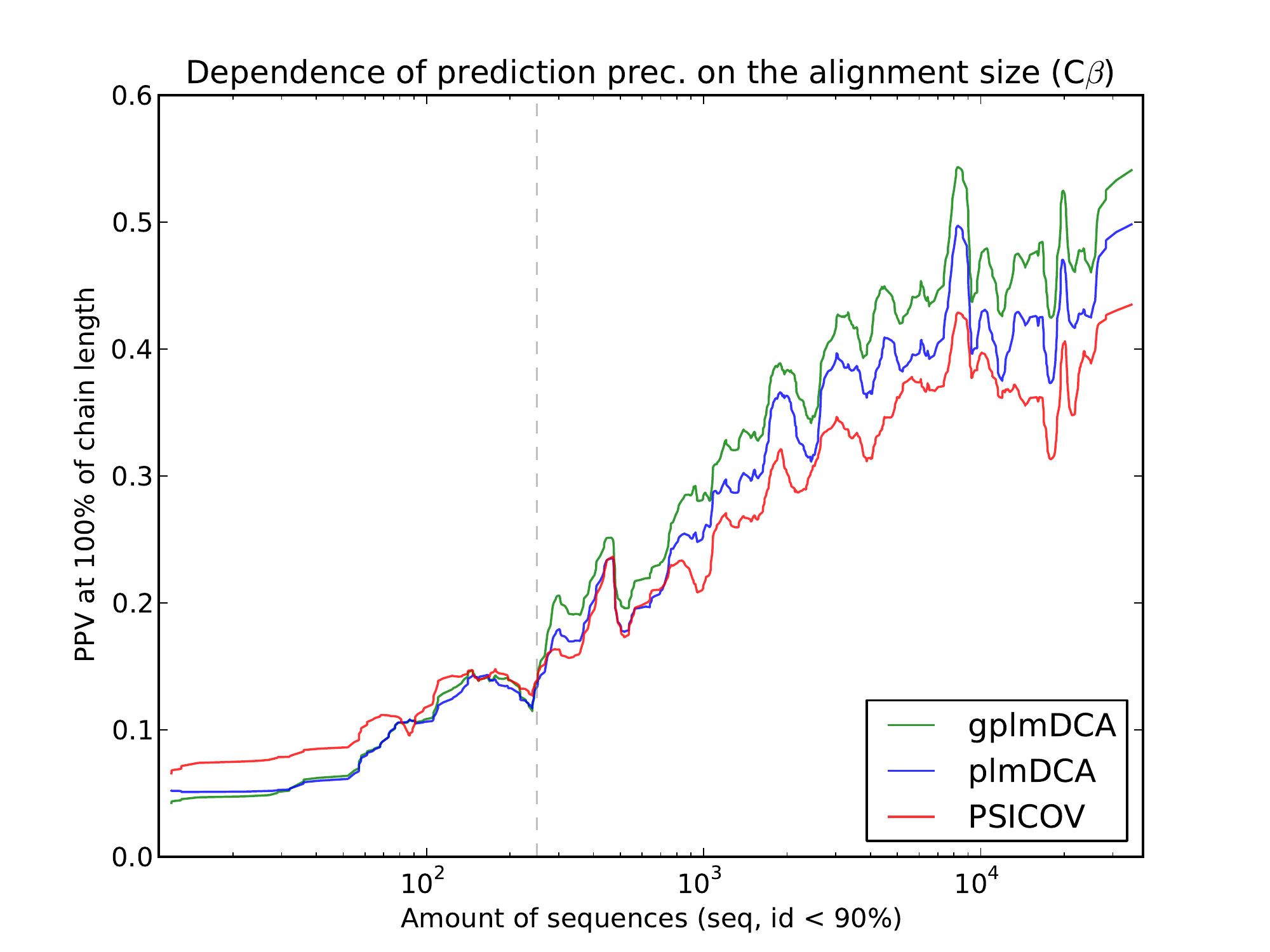}
\end{center}
\caption{
Contact prediction accuracy for proteins in the test set by gplmDCA and plmDCA \textit{vs} number of homology reduced sequences in the alignment (maximum 90\% sequence identity), when considering top 10\%, 25\% (top row), 50\% and 100\% (bottom row) contacts,
100\% being the same number of contacts as the number of amino acids in the protein.
The advantage of gplmDCA is particularly interesting in ranges highlighted by vertical
dotted lines. For the top 10\% and top 25\% (top row) these ranges
are approximately 60-2500 and 250-23000 sequences
(414 and 622 out of 801 proteins), while for the top
50\% and top 100\% (bottom row) they extend from about 250 sequences in the alignment
and upwards (651 out of 801 proteins).
PSICOV outperforms both plmDCA and gplmDCA when there are less than about
100 sequences in the alignment. The peak around
500-sequence point is due to concentration of $\beta$-sheet rich proteins (mostly
hydrolases), that seem to be particularly amiable to contact prediction.}
\label{few-sequences} 
\end{figure}

\paragraph{Adding gap parameters to the model leads to improved predictions when there are few sequences.}
As shown in figure~\ref{few-sequences} the improvement in prediction performance
by using gplmDCA depends on how many sequences there are in an alignment.
When considering the top ranked $\frac{1}{10} \cdot L$ contacts
per protein, where $L$ is protein length, the improvement is centered in an interesting
intermediate range of approximately 60-2500 sequences with at most 90\% sequence similarity, while gplmDCA and plmDCA are
similar in performance when the number of sequences is less than 60
(where it is poor) or more than 2500 (where it has saturated at a PPV around 65\%).
Even with as few as 300 unique sequences in alignment, gplmDCA is able to achieve 40\%
positive prediction rate for these highest ranked contacts. As more contacts are
considered, the range where gplmDCA holds an advantage moves successively to
proteins with more sequences.
A proposed explanation of these observations is that the less information (sequences) are available, the more prominent
the confounding factor of the gaps become for plmDCA. Introducing gap parameters alleviates this phenomenon,
and increases the prediction precision for top ranked contacts for
information-poor alignments and improves the amount of correct contacts predicted for the
information-rich alignments.

%% file: discussion.tex
\section*{Discussion}
While the set of proteins reported in this work is significantly more
''difficult'' than the proteins reported in recent work on the
subject, it is evident that extending the model with a gap term
significantly increases the accuracy of prediction.
This improvement can be attributed to incremental developments in three
aspects, which we call the \emph{three dimensions of contact
prediction}: data, model and method. While each of these aspects has been shown
to have a non-negligible impact on the accuracy of contact prediction on its own, 
this work suggests they should not be considered separately, but rather in unison.

\begin{figure}
\begin{center}
\includegraphics[width=0.68\textwidth]{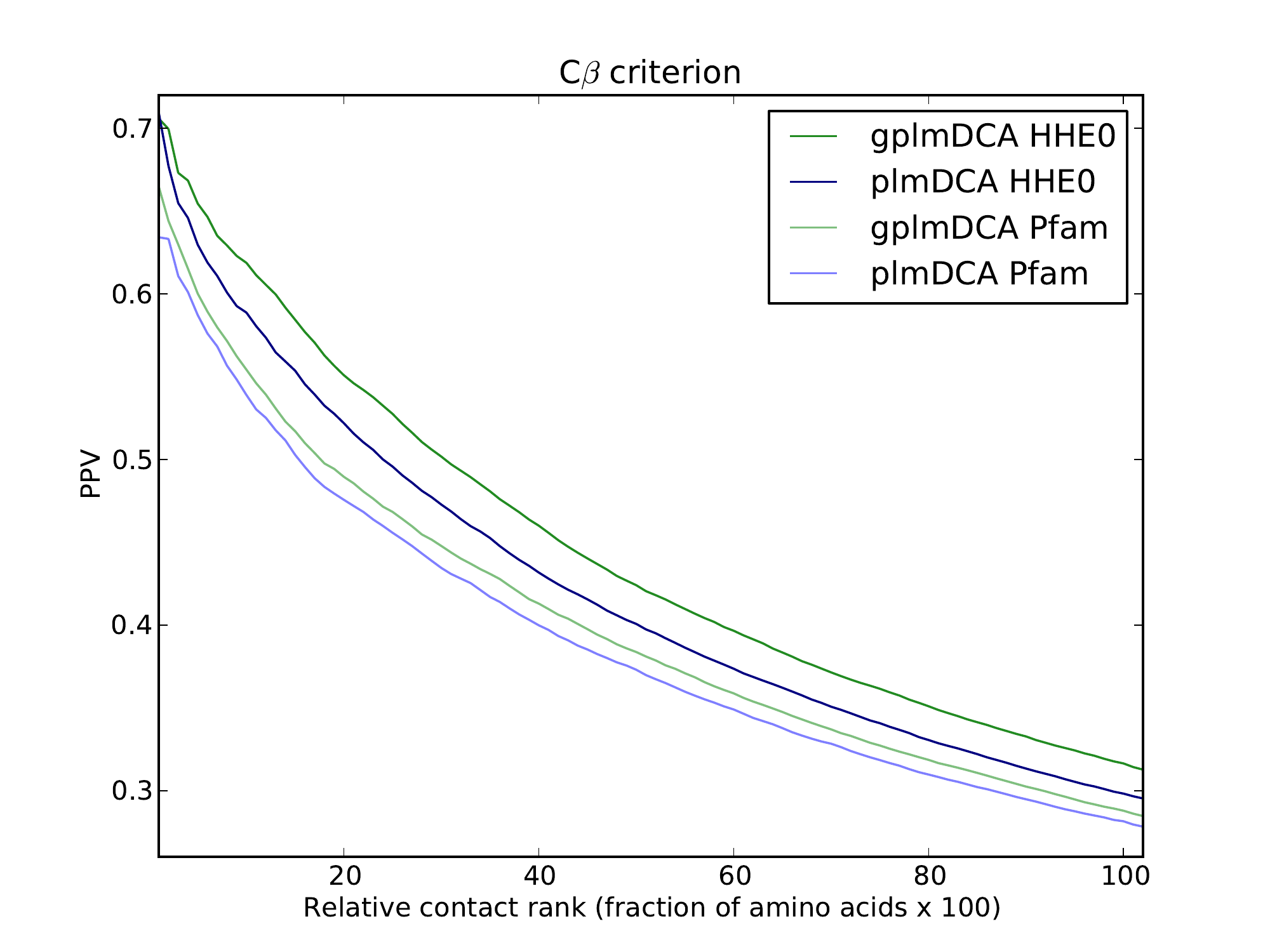}
\end{center}
\caption{Prediction by absolute PPV and $C\beta$ criterion for gplmDCA and plmDCA run on Pfam and HHblits alignments
in the reduced test data set. The reduced test data set comprises the proteins in the main test data set where
a comparison can be made to Pfam alignments, as described in Methods.}
\label{figure-pfam}
\end{figure}

\paragraph{The data.} The extensive benchmark performed for the purposes of the paper has validated
our previous claim that proper input alignment matters for accurate contact
prediction~\cite{skwark2013pconsc}. 
To compare HHblits and Pfam alignments we have from our main data set constructed a \textit{reduced data set}.
As shown in Figure~\ref{figure-pfam} gplmDCA has a larger advantage over plmDCA on HHblits alignments
than on Pfam alignments. Note that plmDCA on HHblits alignments is actually clearly better
than gplmDCA on Pfam alignments, confirming again the importance of the data dimension
in contact prediction.

\begin{figure}
\begin{center}

\includegraphics[width=0.38\textwidth]{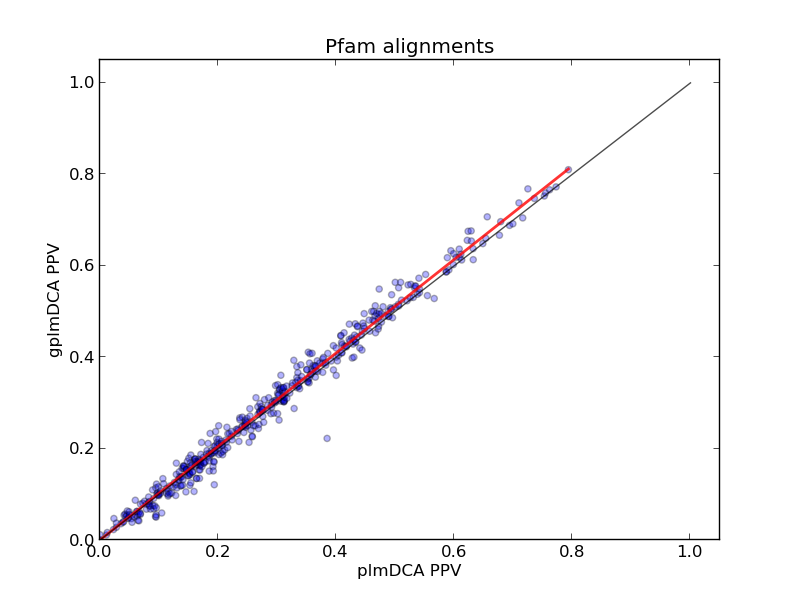} %
\hfill
\includegraphics[width=0.38\textwidth]{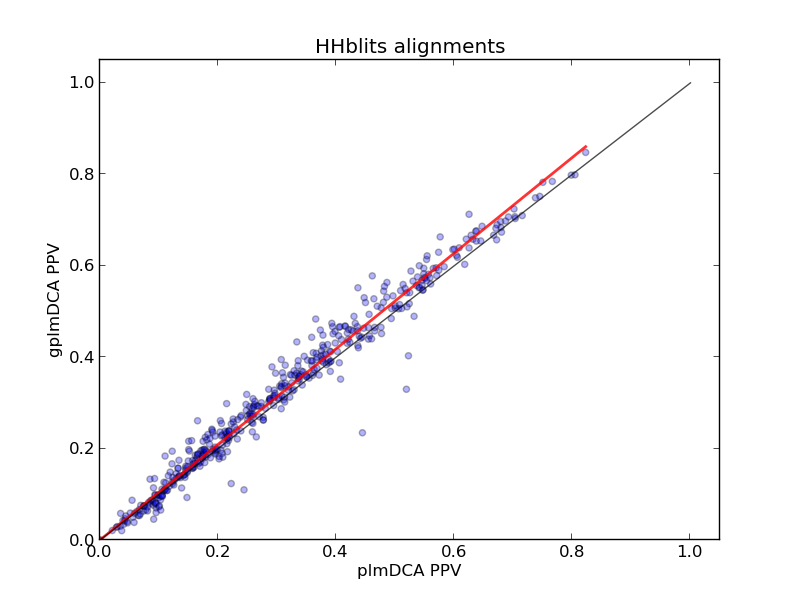} %

\includegraphics[width=0.38\textwidth]{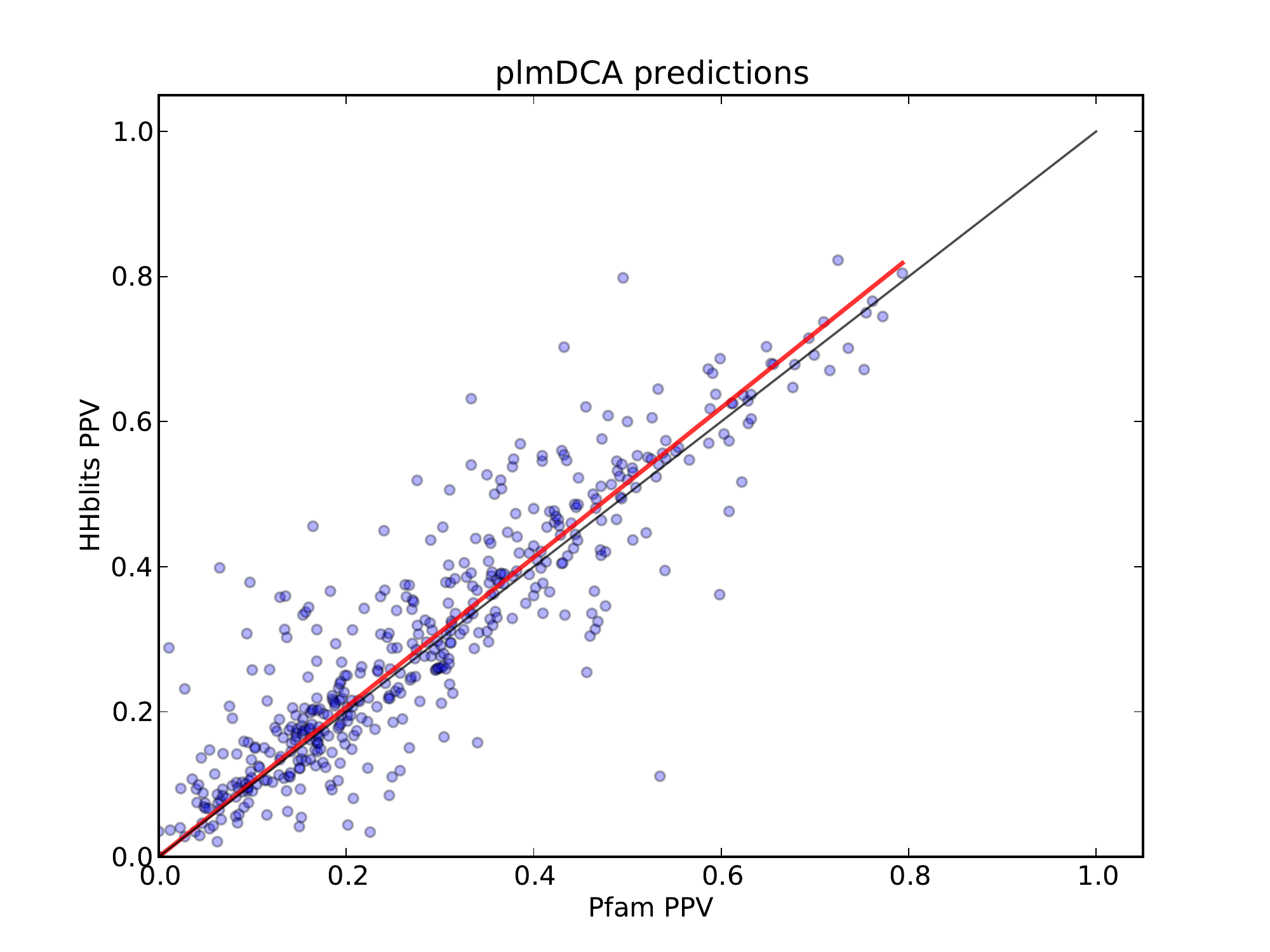} %
\hfill
\includegraphics[width=0.38\textwidth]{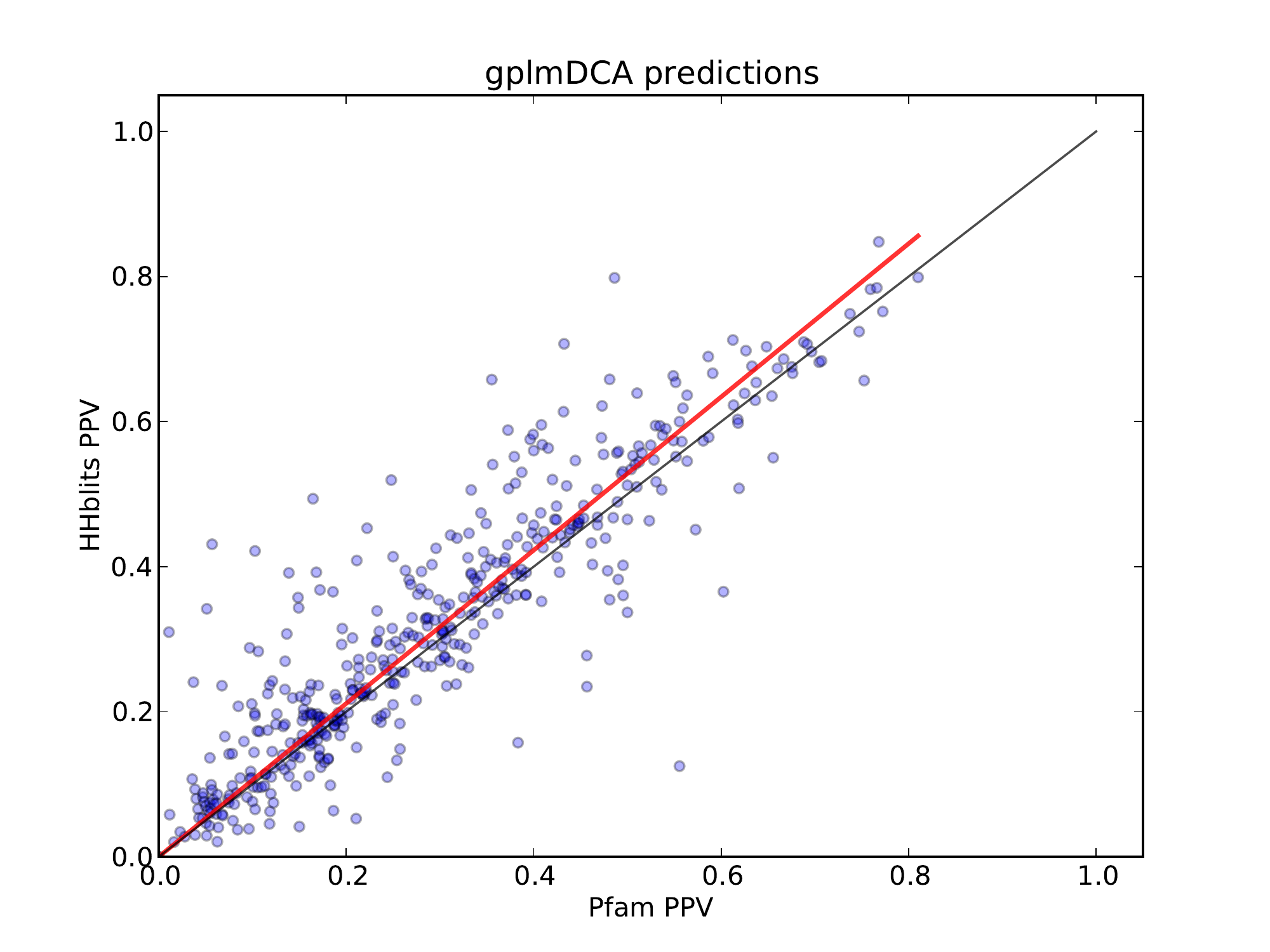} %
\end{center}
\caption{Scatter plots of prediction by absolute PPV and $C\beta$ criterion for individual proteins in the
reduced test data set. Top row shows, analogously to Figure~\protect\ref{scatter-plot} (in Results, for the main data set),
gplmDCA vs plmDCA for Pfam alignments (left panel) and for HHblits alignments (right panel). 
Bottom row shows prediction for HHblits alignments vs Pfam alignments
using plmDCA (left panel) and gplmDCA (right panel).}
\label{figure-pfam-scatter}
\end{figure}

On the level of single proteins, both with Pfam alignments and HHblits alignments, gplmDCA has a clear advantage
over plmDCA in terms of prediction precision, see top row of Figure~\ref{figure-pfam-scatter}. The difference is more pronounced for HHblits alignments, which can 
be quantified by the slope of OLS regression line, that is $1.046 \pm 0.004$ in case of HHblits alignments, but only $1.023 \pm 0.003$ for Pfam alignments.
In the other dimension of the same test, gplmDCA gains more from use of HHblits over Pfam than plmDCA (bottom row of  Figure~\ref{figure-pfam-scatter}),
with the regression line slopes of $1.056 \pm 0.011$ for gplmDCA, and $1.031 \pm 0.011$ for plmDCA.

\paragraph{The model.} Contact prediction in DCA has hitherto been considered in
terms of a pairwise interaction model, typically motivated by maxentropy arguments \textit{cf}~\cite{hopf2012}.
In a context where one tries to learn from all of the data and not from a reduced
set of observables such as \textit{e.g.} pair-wise correlation functions, maxentropy arguments 
do not apply, and there is a vast array of
possible models that could describe the biological reality more accurately.
We have shown here that the addition of what is arguably the simplest 
and most obvious non-pairwise term, the gap term, does make a significant
difference to the quality of resulting contact predictions. Therefore we
strongly posit that \emph{the pairwise interaction term is not the end of the
story, but rather a prelude}, and that there remains a lot that can still be
done in respect to constructing data models that more accurately reflect the
evolutionary relationships in proteins.

\paragraph{More accurate contact maps.} The improvement in terms of average PPV over the whole protein set, as well as
the fraction of proteins for which gplmDCA produces more accurate predictions,
cannot be be underestimated, but is not the only distinguishing feature of
gplmDCA. Eliminating strong couplings induced by gaps in the alignments allows
for detection of relatively weaker ones, which may be important for the future
applications of the method, such as contact-assisted protein folding.

\begin{figure}
\begin{center}
\includegraphics[width=0.58\textwidth]{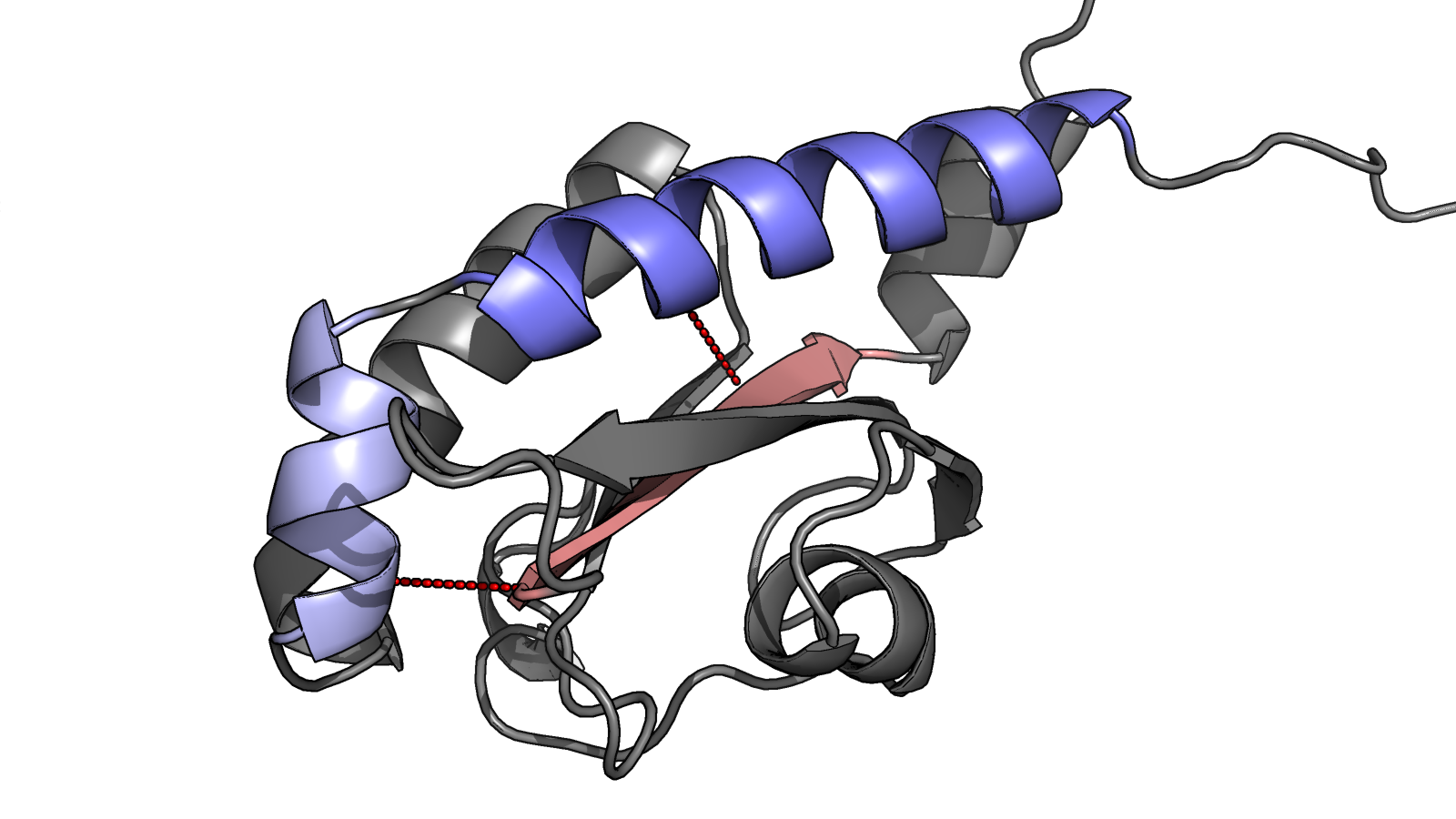}
\includegraphics[width=0.38\textwidth]{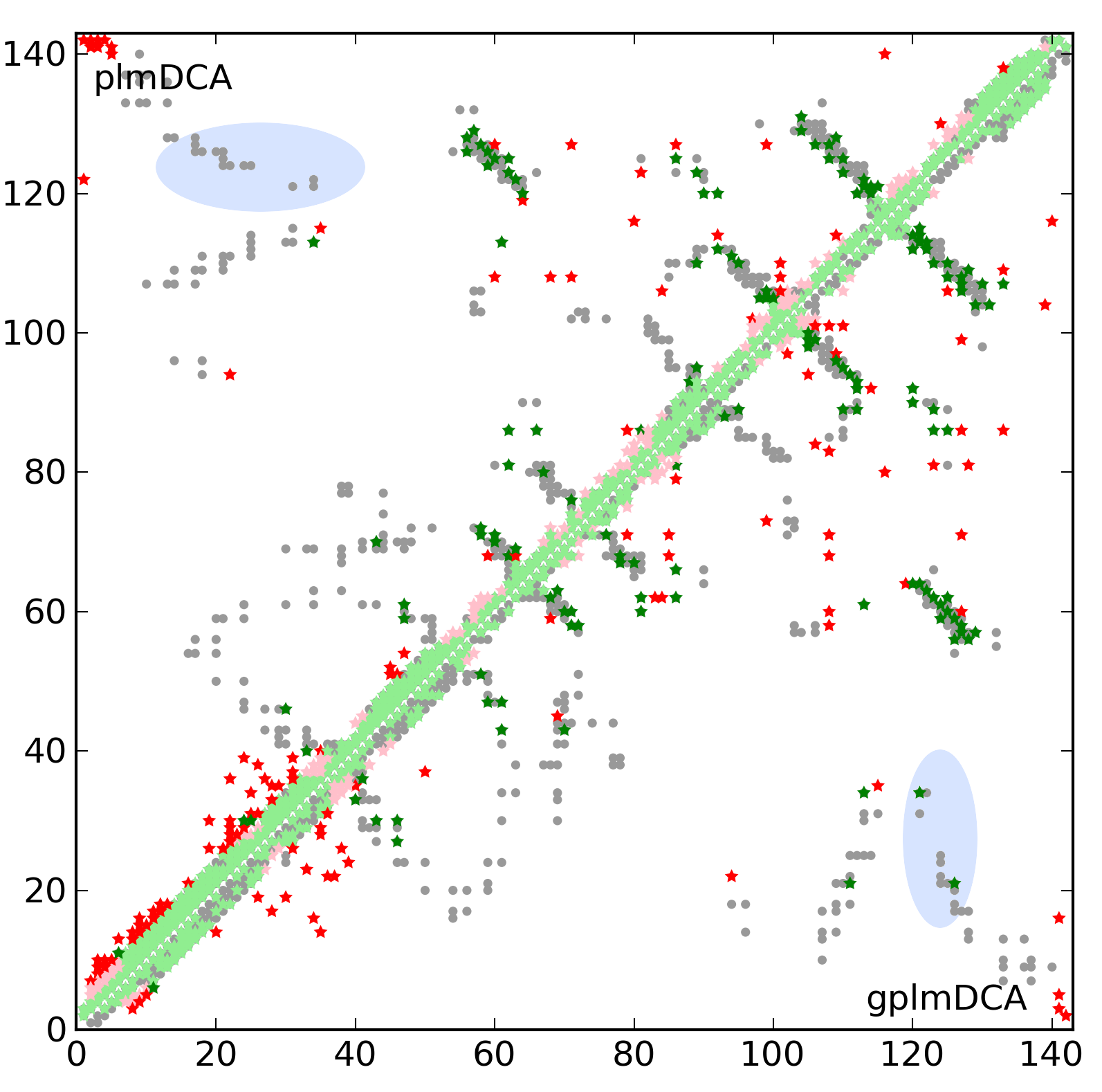}
\end{center}
\caption{Difference in contact prediction between plmDCA and gplmDCA for sensor domain of histidine kinase DcuS from E.coli (pdbid:3by8\_A). Left figure: protein structure, with some of contacts uniquely predicted by gplmDCA marked by dashed lines. Right: contact map, with the region of interest marked in faint blue.}
\label{fig-3by8}
\end{figure}

One example of such contacts being predicted, shown in Figure~\ref{fig-3by8}, is the contacts between
N-terminal helices (marked in blue) and the $\beta$-sheet of the sensor domain
of histidine kinase DcuS (deposited in PDB as 3BY8\_A). This structure is classified in CATH~\cite{orengo1997cath} as a
two-layer sandwich and while plmDCA is able to position strands of the
$\beta$-sheet in a correct order, it fails at predicting contacts between the
$\alpha$-helices of the sandwich and the $\beta$-sheet. As can be seen in
Figure~\ref{fig-3by8}, gplmDCA in addition to the already predicted contact
between residues 34 and 113 (green star next to the blue region) predicts also
contacts between residues 34 and 121, as well as 21 and 126. This in theory
should allow for proper positioning of helices in case of strucutre prediction.

\paragraph{Wrong predictions.} The addition of a gap term, while beneficial for
vast fraction of proteins, occasionally results in lower prediction accuracy in
comparison to the inference performed on a model without gap term (plmDCA).

\begin{figure}
\begin{center}
\includegraphics[width=0.48\textwidth]{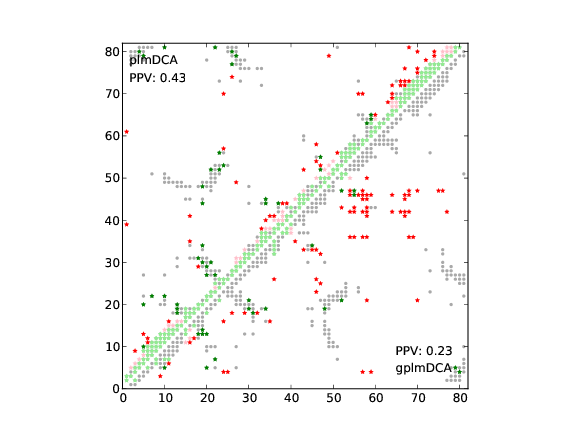}
\includegraphics[width=0.48\textwidth]{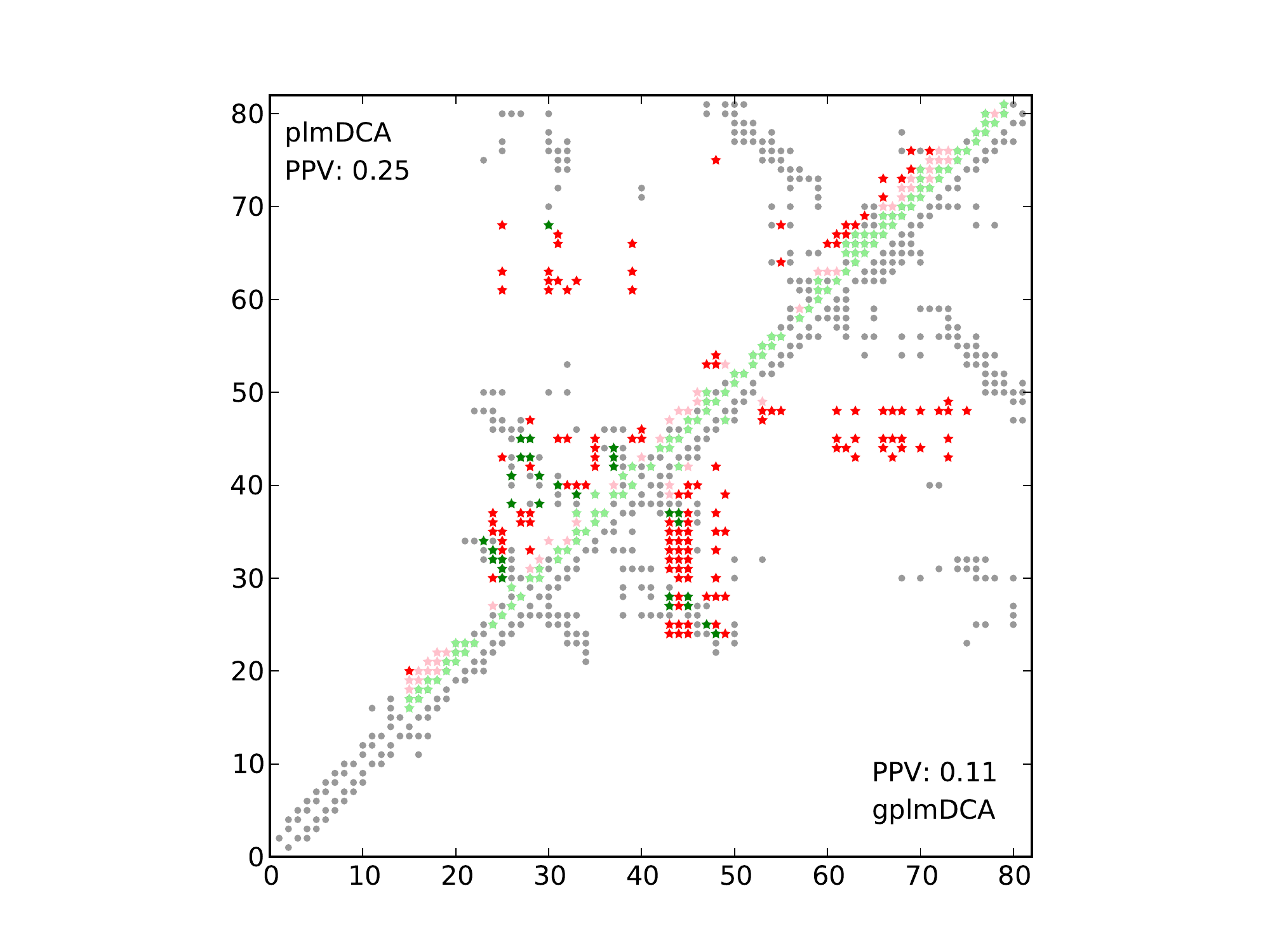}
\end{center}
\caption{Mispredictions. Among the 801 proteins
plotted in Figure~\protect\ref{scatter-plot} there are two prominent outliers
where plmDCA (model with no gap parameters) clearly does better than
gplmDCA (model with gap parameters). Left panel shows the contact map
of b558 where gplmDCA wrongly predicts a number of contacts between 
amino acids 35-45 and 55-70. Right panel, contact map of Spt4. For discussion,
see main text.}
\label{misprediction}
\end{figure}

A most striking example is soluble cytochrome b558 (a prokariotic homolog of
cytochrome b5) from Ectothiorhodospira vacuolata (deposited in PDB as 1CXY:A),
which is the most prominent outlier in Figure~\ref{scatter-plot}.
While plmDCA predicts contacts allowing for proper assembly of protein (at
least in the $\beta$-sheet region), gplmDCA predicts significantly fewer such
contacts, but -- more importantly -- neglects to predict nearly all close range
contacts. We have found nothing immediately obvious that would make the multiple sequence
alignment constructed for this protein unsuitable to contact prediction. The alignment 
has nearly 7000 homologous, appropriately diverse protein sequences, with
proper coverage across the whole span of the protein chain. We note 
that predictions conducted on a slightly thinner alignment (including homologs with e-value
cutoff of $10^{-4}$, instead of $10^0$, resulting in ~300 fewer sequences), or alignments of similar size
produced by different methods (i.e. jackhmmer), do not seem to exhibit such a behavior.

The other outlier in Figure~\ref{scatter-plot}, albeit less prominent, is transcription elongation factor Spt4 from Pyroccocus furiosus (deposited in PDB as 3P8B:A). 
In this case, all the contacts predicted by gplmDCA concentrate in rectangular regions between residues 24-49, 53-56, 59-75, which we believe
could be due to the high percentage of sequences with identical gap distribution in the alignment, either
(case 1) 1-23, 50-52, 56-59, 77-81 (31.7\% of sequences) 
or (case 2) 1-23, 50-52, 56-59, 64-65, 74-81 (28.4\% of sequences). 
\paragraph{Folding.} Elimination of artifacts in predicted contact maps, as
well as increased sensitivity (predicting correct contacts between more
secondary elements) in comparison to plmDCA, coupled with increased prediction
precision, strongly suggest that gplmDCA should provide valuable input for the
future \textit{ab-initio} protein structure prediction attempts. The previous
incarnation of pseudolikelihood maximization for direct coupling analysis
(plmDCA) has been succesfully used for protein structure prediction endeavors
(c.g.~\cite{Marks2012-NatureBio}) as it objectively provides higher prediction accuracy
than other methods (as demonstrated, for example in~\cite{skwark2013pconsc}).
As gplmDCA is both faster and more accurate than the version used in reported
structure prediction work we strongly recommend it for future use.

\paragraph{Conclusion.} 
Contact prediction has advanced greatly in the last five years, reaching
a level of accuracy which was previously believed to be unattainable.
We have shown here that the three \emph{dimensions} of data, model and method
are all important for overall prediction success, and we have
shown that one can can significantly improve prediction along the second
dimension by going beyond pairwise maxentropy models mainly used in the field
up to now.  We believe that these are only the first steps in a rational
approach to incrementally improve contact prediction, and that with the ongoing
explosion in the number of available protein sequences much further progress
should be possible on these issues.

%% file: methods.tex
\section*{Methods}

The Direct Contact Analysis (DCA) as introduced
in~\cite{lapedes_correlated_1999} and~\cite{weigt_identification_2009} is a
family of methods to predict contact between amino acid pairs from a multiple
sequence alignment
(MSA)~\cite{morcos_direct-coupling_2011,marks_protein_2011,balakrishnan2011,jones2012,hopf2012,ekeberg_improved_2013,skwark2013pconsc,cocco_principal_2013,Burkoff2013,Lui2013,Rivoire2013,Andreatta2013}.
Learning predictive models of amino acid contacts depends on which sequences
are used to build the alignment and by which methods they are aligned (Input
data), which model one tries to learn from the data (Model) and how a model is
learnt from the data (Inference method).  We describe below our approach along
these three dimensions in turn.  The perceived quality of prediction then depends on how
the model is used and how it is benchmarked, as we describe below (Prediction
and benchmarking metrics).

\paragraph{Input data}
In a substantial fraction of the contributions to the development of DCA
contact predictions have been based on MSAs obtained from the Pfam protein families database:~\cite{punta2012,pfamsite}.
However, as recently shown by one of us in~\cite{skwark2013pconsc}, and as also shown here (see Discussion),
these alignments are not the optimal input for DCA and DCA-like methods.

Instead of PfamA alignments, we use a state-of-art homology detection
method HHblits~\cite{soding2012}, based on iterative comparison of Hidden Markov models (HMMs).
This approach is able to arrive at very accurate
mulitple sequence alignments, tailored to the protein of interest, while still
including remotely homologous proteins. 

We have constructed a heterogenous set of 801 protein chains of known
structure, sampled from Protein Data Bank which we refer to as \textit{main test set}. 
This set is an amalgam of four smaller data sets as follows:
\begin{itemize}
\item 150 proteins reported in PSICOV paper~\cite{jones2012}
\item $\sim$ 150 proteins with known structures, with relatively few detectable homologous proteins of known sequence.
\item $\sim$ 200 proteins of the most common Structural Classification of Proteins (SCOP) folds~\cite{scop95}
\item $\sim$ 300 proteins sampled at random from PDB
\end{itemize}

We excluded from the main test set proteins that were significantly too
long for a reasonable contact prediction (the mean and median lengths of a
protein in the considered set are 168.4 and 149 amino acids corresponingly,
with maximum of 404 amino acids), or not compact enough (probably stabilized by
interaction with their environment).  We did not exclude
multimeric proteins, or filter out multidomain proteins, though.

The alignments in the main test set 
have been constructed using HHblits, as contained in HHsuite 2.0.16 with a bundled uniprot20\_2013\_03 database. 
We have run five iterations of search, with a E-value cutoff of 1, allowing for
inclusion of distantly homologous protein in the alignment. The search was
conducted without filtering the result MSA (-all parameter), without limiting
the amount of sequences allowed to pass the second prefilter and allowing for
realigning all the hits, hence obtaining the most information-rich and accurate
alignment at cost of increased running time. 

To compare Pfam and HHblits-based predictions we have from the main test set also constructed
a \textit{reduced test set} by the following procedure. For each of the proteins in
the main test set we searched for its PDB identifier against an official Pfam-PDB
mapping, to identify the longest Pfam family corresponding to this protein (in
case of potential multiple Pfam hits per PDB identifier).
This resulted in alignments for 481 proteins, reflecting \textit{inter alia} the fact that not all 
proteins in the main test set have an official Pfam-PDB mapping.
Then we identified
the sequence in the appropriate Pfam alignment which is closest to the sequence
of protein in question by Smith-Waterman algorithm using BLOSUM100 matrix.  
From this set we reject alignments where we the number of residues in both sequences aligned to gaps
is more than 50\% of length shorter of sequences plus length difference between
sequences, and subsequently we trim the Pfam alignment to only the columns aligned
to protein in question. Finally, the reduced test set contains 451 proteins
with both Pfam and HHblits MSAs which form the input plmDCA and gplmDCA in
the comparisions presented in Discussion and Figure~\ref{figure-pfam}.
The comparison is there done by filtering down the predictions to include only the columns present in the Pfam alignments.  

Protein sequences present in sequence database (and hence used for alignments in this work) are biased towards sequences from genomes of
organisms that are of special interest to humans.  Many such sequences are
closely similar, and following~\cite{weigt_identification_2009} sequences that
are more similar than some threshold are reweighted before being used in a DCA.
We here use the reweighting recently described in~\cite{ekeberg2014}, with
threshold $0.1$, that is, by reweighting sequences that are more than 90\% identical.

\paragraph{Model} 
A multiple sequence alignment can be considered
as samples from an unknown probability distribution. Each row,
corresponding to one protein in the alignment, is then one
of the $q^N$ possible realizations of a random variable which at each
of the $N$ positions along the row can take $q=21$ different values 
(the amino acid or the gap symbol at that position). The (unknown) probability 
distribution is, in principle, the result of the complete evolutionary history
of all forms of life, and is therefore a very complicated object. However,
it is not necessary to know the probability distribution exactly to extract useful
information. 

The Direct-Coupling Analysis (DCA), as introduced in~\cite{lapedes_correlated_1999}
and~\cite{weigt_identification_2009},
assumes that the probability distribution is the \emph{Potts Model} of statistical physics~\cite{wu1982potts}:
\begin{equation}
P_{Potts}(\underline{a}) = \frac{e^{-H_{Potts}(\underline{a})}}{\mathcal{Z}} \qquad H_{Potts}(\underline{a}) = -\sum_{i<j} J_{ij}(a_i,a_j) - \sum_{i} h_i(a_i).
\label{eq:Potts}
\end{equation}
The use of the Potts model in the DCA has often been motivated by \emph{maxentropy} arguments \textit{cf}~\cite{hopf2012}.
As we base our approach an inference method which uses all the data (see below), 
we cannot refer to \emph{maxentropy} principles. Instead, one may observe that 
it has been found in many branches of science and engineering, that 
probability distributions over a collection of a large number of similar objects 
often obey a large deviation principle~\cite{Varadhan1984}. The full distribution
$P$ can then be written as $P(\underline{a}) \approx \exp\left(-L(\underline{a})\right)$, where the
function $L$ in the exponent is ``simple'', a classical example being the Gibbs-Boltzmann distribution
of equilibrium statistical mechanics. An unknown probability distribution can then 
be expanded in a series 
\begin{equation}
-\log P(\underline{a}) = L(\underline{a})=\hbox{Constant} + S_1(\underline{a}) + S_2(\underline{a}) + \ldots
\label{eq:Amari-expansion}
\end{equation}  
where the first order contribution $S_1$ (linear) contains terms only depending on one component of $\underline{a}$, 
the second order contribution $S_2$ (bi-linear) contains terms depending on two components of $\underline{a}$, and so on. 
If $L$ in fact \textit{is} simple, then a low order truncation should give a useful approximation to $P$,
and the Potts model of (\ref{eq:Potts}) is nothing but the truncation of (\ref{eq:Amari-expansion}) after the second order terms.
We note that hierarchies of exponential probability distributions have non-obvious properties, and may for instance
be taken as a basis of an invariant decomposition of the entropy~\cite{Amari2001}.

Any multiple sequence alignment procedure typically produces stretches of gaps,
a fact which is obvious by visual inspection. It is therefore an immediate observation 
that a real MSA data cannot be a set of independent realizations of the rather simple model in (\ref{eq:Potts}),
since such stretches of one and the same variable (the gap variable) are very unlikely to occur in 
a random variable drawn from the distribution (\ref{eq:Potts}).
In a DCA based on (\ref{eq:Potts})
we manifestly learn from data a model which does not generate the same data. 
We therefore hypothesized that by learning a model which describes the data
better, we might also better predict amino acid contacts. 

To investigate this we introduced additional gap parameters and try to learn
\begin{equation}
P_{Gap-Potts}(\underline{a}) = \frac{e^{-H_{Potts}(\underline{a})-H_{Gap}(\underline{a})}}{\mathcal{Z}}\quad
H_{Gap}(\underline{a}) =  - \sum_{l=1}^{L} \sum\limits_{i=1}^{N-m+1} \xi_{i}^{l} I_i^{l}(\underline{a}),
\label{eq:gapHamiltonian}
\end{equation}
where the $\xi_{i}^{l}$ are new parameters describing the propensity of a
site $i$ to be the beginning of a gap of length $l$,
$I_i^{l}(\underline{a})$ is an indicator function which takes the value $1$ if
there is a gap of length $l$ beginning at site $i$, and otherwise zero, and 
$L$ is a meta-parameter, the largest gap length included in
the gap parameters. We set $L$ to the largest gap length found in a
given alignment. 
The number of additional parameters to be learned is thus not larger than $NL$,
to be compared to the number of parameters already used in (\ref{eq:Potts}), which is about $\frac{1}{2}q^2N^2$.

\paragraph{Inference Method} 
The benchmark of learning a model from data is maximum likelihood where one chooses the
probability distribution in
a class which minimizes a negative-log-likelihood function $L$. The main problem in learning
(\ref{eq:Potts}) from data
by maximum likelihood is that the normalizing constant ($\mathcal{Z}$) cannot be evaluated 
exactly and efficiently in large systems, and that therefore maximum likelihood learning 
can only be done approximately \textit{e.g.} by variational methods~\cite{WainwrightJordan}.
Therefore, we instead use the weaker learning 
criterion of pseudo-likelihood maximization~\cite{besag1975}, 
first applied in the DCA setting by one of us in~\cite{ekeberg_improved_2013}.
A further issue is that the number of parameters in a Potts model based DCA 
is (typically) larger than the number of observations (number of sequences in
an MSA), and regularization is therefore necessary.
We here base our work on the recently developed \textit{asymmetric pseudo-likelihood maximization}~\cite{ekeberg2014},
which is considerably faster than the version presented in~\cite{ekeberg_improved_2013} 
while showing essential identical performance as a predictor of amino acid contacts.

Learning the new model including (\ref{eq:gapHamiltonian}) is especially convenient
using the pseudo-likelihood maximization approach. We have developed a new
code gplmDCA based on the asymmetric version of plmDCA of~\cite{ekeberg2014}.
Regularization is by an $L_2$ norm on parameters as described in~\cite{ekeberg2014}.

\paragraph{Prediction and benchmarking metrics}. 
The outcome of learning a model of the Potts type is a set of pairwise
interaction coefficients $J_{ij}(a_i,a_j)$. For each pair $(i,j)$ (each pair
of positions) this is a matrix in two other variables ($a_i$ and $a_j$)
and how an inferred interaction is scored depends on which matrix norm one
uses. We here use the Frobenius norm augmented by the Average
Product Correction (APC), as introduced in the context of DCA by one of 
us in~\cite{ekeberg_improved_2013}, and order the pairs $(i,j)$, for each multiple sequence alignment, by the value of this score.
 
To benchmark the predictions of the DCA one compares against known crystal structures.
In this work we use as the main benchmark criterion, that two
amino acids are in contact, if their $C\beta$ atoms are at most $8\hbox{\AA}$
apart in the crystal structure. This we denote as \emph{$C\beta$ criterion} and
use predominantly throughout this article. In order to facilitate 
comparison to previously published work on the DCA we 
present also an alternate metric that considers the amino acids to be in
contact if any of their heavy (non-hydrogen) atoms are at most $8.5\hbox{\AA}$
apart. This metric is denoted as \emph{$8.5 \hbox{\AA}$ heavy atom criterion} and discussed in the supplementary material.

In this article we use the terms \emph{precision} and \emph{PPV} (positive predictive value)
interchangably, with metric denoting the ratio of true positives
to all predictions (within a certain count threshold). In line with previously
published work on contact prediction, we consider only the contacts with
sequence separation greater or equal to 5 amino acids (we do not consider very
short range contacts, that is contacts between amino acids $i$ and $j$ when $|i-j| < 5$). 

By the term \emph{weighted moving average} with window $w$, authors understand
a weighted arithmetic mean of a value at a given position and $w$ values on
either side of the center position, thus resulting in $2 \cdot w + 1$ values to
be averaged. The central position is scaled with weight $w$, whereas the
weights decrease in aritmetic progression while moving away from the center
(i.e positions $-1$ and $+1$ are scaled with weight $w-1$, whereas positions $-2$ and $2$ with weight
$w-2$ etc.).

\paragraph{Availability.} The code of gplmDCA is freely
available at \texttt{http://gplmdca.aurell.org}. This website contains also
a link to all the data the benchmark is based on, that is: multiple sequence
alignments, predicted couplings (both plmDCA and gplmDCA), protein structures
and contacts derived from them.